\documentstyle[aps,psfig]{revtex}

\begin{document}
\draft 

\twocolumn[\hsize\textwidth\columnwidth\hsize\csname@twocolumnfalse\endcsname

\title{ The first principles calculation of transport coefficients }

\author{ Dario Alf\`e}
 
\address{ Geological Science Dept., University College London,
Gower Street, London WC1E 6BT, U.K.}

\author{ Michael J. Gillan  }

\address{ Physics Department, Keele University, Keele, Staffordshire
ST5 5BG, U.K. }

\date{\today} \maketitle

\begin{abstract} 

We demonstrate the practical feasibility of calculating transport
coefficients such as the viscosity of liquids completely from first
principles using the Green-Kubo relations. Results presented for
liquid aluminum are shown to have a statistical error of only ca. $
5 \%$. The importance of such calculations is illustrated by results
for a liquid iron-sulfur alloy under Earth's core conditions, which
indicate that the viscosity of the liquid outer core is not
substantially higher than that of typical liquid metals under ambient
conditions.

\end{abstract}

]

\narrowtext

Since the pioneering work of Car and Parrinello \cite{cp} first
principles molecular dynamics (FPMD) has become a widely used
technique for investigating condensed matter. For liquids and solids
in thermal equilibrium many quantities of interest are readily
calculated, including thermodynamic functions, structure factors,
radial distribution functions, diffusion coefficients, bond lifetimes,
etc. But transport coefficients such as viscosity and thermal
conductivity are computationally more demanding, and have been little
studied by FPMD.  We demonstrate here the feasibility of
calculating such quantities with useful accuracy, by presenting FPMD
results for the viscosity of liquid aluminum near the triple point.
We then illustrate the potential importance of this type of
calculation by showing how FPMD calculations can be used to calculate
the viscosity of liquid iron in the Earth's outer core. This is one of
the key quantities in the theory of the Earth's deep interior, but
also one of the most uncertain.

There are basically two ways of calculating transport coefficients by
simulation. The first is to subject the simulated system to an
explicit external perturbation (e.g. a shear flow or a temperature
gradient) and calculate the steady-state response. Alternatively, one
can apply the Green-Kubo (GK) relations, i.e. the relations between
transport coefficients and correlation functions involving fluxes of
conserved quantities \cite{allen}.  The shear viscosity $\eta$,
for example, is given by:
\begin{equation}\label{shear}
\eta = \frac{V}{k_BT} \int_0^\infty dt \langle P_{xy}(t)
P_{xy}(0) \rangle,
\end{equation}
where $\langle \cdot \rangle$ denotes the thermal average, $V$ is the
volume of the system, $T$ is the temperature, $k_B$ is the Boltzmann
factor, and $P_{xy}$ is the off-diagonal component of the stress
tensor $P_{\alpha\beta}$ ($\alpha$ and $\beta$ are Cartesian
components).  The relative merits of the two approaches have been
much discussed, but the GK method has the virtues of simplicity and
ease of application, and will be used here.

For a simulation having a given duration, single particle properties,
like the diffusion coefficient, can be calculated more accurately than
collective properties, like the viscosity. In the former case the
statistical average can be done over time and over the particles,
while in the latter one loses the possibility of averaging over the
particles. To obtain the same statistical accuracy, collective
properties need much longer runs than single particle properties by a
factor proportional to the size of the system.

The first principles calculations presented here are based on density
functional theory, pseudopotentials and plane-wave basis sets.  The
electron-ion interaction is described by ultrasoft Vanderbilt
pseudopotentials (PP)\cite{vanderbilt}.  Non-linear core corrections
\cite{louie} are included throughout this work. The calculations have
been performed using VASP (Vienna ab initio simulation
package)\cite{kresse}. In VASP the electronic ground state is
calculated exactly (within a self-consistency threshold) at each MD
step, using an efficient iterative matrix diagonalization scheme and a
Pulay mixer \cite{pulay}.  Within this approach to FPMD it is
important to provide a good starting electronic charge density at each
time step, so as to reduce the number of iterations to achieve
self-consistency. We have used the following charge extrapolation
scheme: at the beginning of each time step the electronic charge
density is extrapolated using the atomic charge density and a
quadratic extrapolation on the difference, i.e. the charge is written
as: $\rho(t) = \rho_{at}(t) + \delta \rho(t)$ where $\rho(t)$ is the
self-consistent charge density at time $t$, and $\rho_{at}(t)$ is the
sum of the atomic charges. At time $t+dt$ the charge is written as the
sum of the atomic charges, which can be calculated exactly and
cheaply, and a quadratic extrapolation on $\delta \rho$. This scheme
provides a much better starting charge than the standard quadratic
extrapolation of the whole charge \cite{prb}. The electronic levels
are occupied according to Fermi statistics corresponding to the
temperature of the simulation. This prescription also avoids problems
with level crossing during the self-consistent cycles.  Forces and
stress tensor are calculated using the Hellmann-Feynman theorem.  The
temperature is controlled using a Nos\'e thermostat
\cite{nose}.

Our calculation of the viscosity of liquid aluminum was performed
using the local density approximation (LDA) \cite{lda} for the
exchange-correlation energy and a plane-wave cutoff of 130 eV.  The
temperature was 1000 K (70 K above the melting point) and the pressure
was close to zero. We used a cell containing 64 atoms with periodic
boundary conditions and $\Gamma$ point sampling only.  To integrate
Newton's equation of motion for the ions we used the Verlet
\cite{verlet} algorithm with a time step of 3 fs, and the convergence
threshold on the total energy was $1.5 \times 10^{-8}$ eV/atom; with
these prescriptions the drift in the total energy was $\approx 0.2$
meV/atom per ps.  The length of the simulation was 80 ps.  The
simulation was split into three parts, of lengths $\approx 40$ ps,
$\approx 25$ ps, and $\approx 15$ ps, and at the beginning of each
part the Nos\'e thermostat was reinitialized, so as to avoid possible
problems with the accumulation of the drift in the total energy.  The
cell volume was adjusted to give zero pressure; this resulted in a
density of $\rho=2470$ kg m$^{-3}$, which is about $5\%$ larger than
the experimental one \cite{crc}. This discrepancy is probably due
mostly to the LDA.

There are five independent components of the traceless stress tensor,
$P_{xy}$, $P_{yz}$, $P_{zx}$, $\frac{1}{2}(P_{xx}-P_{yy})$,
$\frac{1}{2}(P_{yy}-P_{zz})$, the last two being equivalent to the
first three by rotational invariance, so one can construct five
independent stress autocorrelation functions (SACF). These are
calculated by averaging $P_{\alpha\beta}(t+t_0)P_{\alpha\beta}(t_0)$
over time origins $t_0$.  Since reinitialization of the thermostat
causes small discontinuities, this averaging is restricted so that
$t_0$ and $t+t_0$ never span a discontinuity.

\begin{figure}
\centerline{\psfig{figure=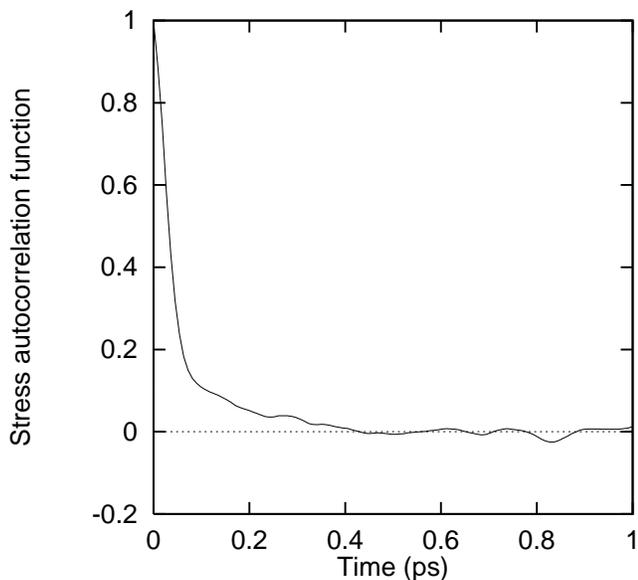,height=3in}}
\caption{Average over the five independent components of the
autocorrelation function of the traceless stress tensor $\phi(t)$
calculated for liquid Al at 1000 K at the density of 2470 kg
m$^{-3}$. Values of $\phi(t)$ are normalized by dividing by
$\phi(0)$.}\label{stressal}
\end{figure}

In Fig. \ref{stressal} we display the average $\phi(t)$ of the five
SACF's divided by its value at $t=0$ for the aluminum system.  Since
the traceless part of the stress tensor has zero average, $\phi(t)$
goes to zero as $t \rightarrow \infty$.  The statistical error on
$\phi(t)$ for all values of $t$ is $\approx 1.5\%$ of the value at
$t=0$ and after $ 0.3-0.4$ ps the magnitude of $\phi(t)$ falls below
that error.

\begin{figure}
\centerline{\psfig{figure=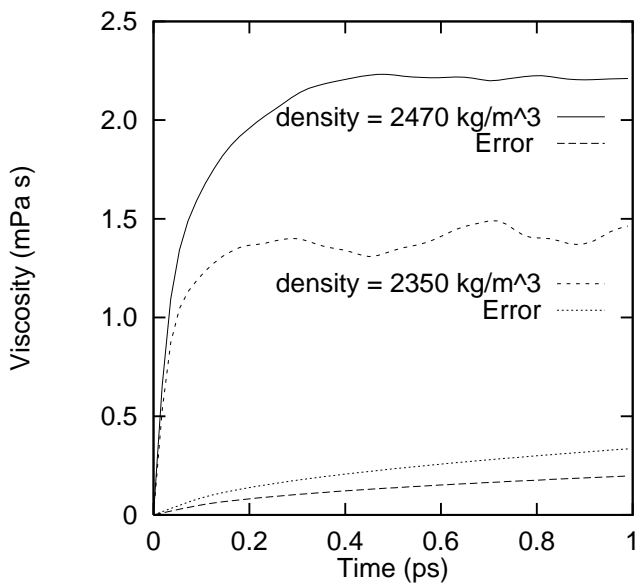,height=3in}}
\caption{Viscosity integral of the average stress autocorrelation
function and its statistical error as a function of time for liquid Al
at 1000 K. Results are shown for the calculated zero pressure density
$\rho=2470$ kg m$^{-1}$ and the experimental density $\rho=2350$ kg
m$^{-1}$.
}\label{viscal}
\end{figure}

In Fig. \ref{viscal} we display the integral $\int_0^t dt'\phi(t')$ of
$\phi(t)$ as a function of time. The limit value of the integral for
$t\rightarrow \infty$ is the shear viscosity. The error that one makes
in evaluating that integral grows with time, since one integrates the
noise together with $\phi(t)$.  We estimated the error in the integral
as a function of time using the scatter of the SACF's calculated by
splitting the simulation into many short intervals. We combined this
estimate with an analytic expression for the error, and the resulting
error estimate is displayed in Fig. \ref{viscal}.  From the point
where $\phi(t)$ falls below the noise one integrates only the latter,
so one gains nothing by evaluating the integral beyond that point.  If
we assume that $\phi(t)$ is zero above $t\approx 0.4$ ps, we obtain
the value $\eta = 2.2 \pm 0.1 $ mPa~s, which should be compared with
the experimental value of $1.25$ mPa~s \cite{shimoi}.  The fact that
the statistical errors have been reduced to $\approx 5\%$ demonstrates
that the viscosity of liquids can be calculated completely from first
principles with quite high precision.  Our calculated value of the
viscosity differs significantly from the experimental one. In order to
find a possible reason for this discrepancy we have simulated for 20
ps the aluminum system at the experimental density, $\rho = 2350$
kg~m$^{-3}$ (which gives a calculated negative pressure of 2 GPa). The
results for the viscosity of this simulation are also reported in
Fig. \ref{viscal}. We obtain $\eta = 1.4 \pm 0.15 $ mPa~s, which is in
excellent agreement with the experimental data.

A method which has sometimes been used in the past to obtain an
approximate estimate of the shear viscosity exploits its
connection with the self-diffusion coefficient $D$ {\it via} the
Stokes-Einstein relation:
$ D\eta = k_B T / 2\pi a, $
where $a$ is an effective atomic diameter.  This relation is exact for
the Brownian motion of a macroscopic particle of diameter $a$ in a
liquid of viscosity $\eta$, but it is only approximate when applied to
atoms. However, if $a$ is chosen to be the radius $r_1$ of the first
peak in the radial distribution function, the relation usually
predicts $\eta$ to within $40 \%$.  We have calculated $D$ from our
simulated liquid aluminum in the standard way from the long time slope
of the mean-square displacement $ \langle \ |{\bf r}_i(t_0+t) - {\bf
r}_i(t_0)|^2 \rangle \rightarrow 6 D t, ~~ {\rm as} ~~ t
\rightarrow \infty,$ where ${\bf r}_i$ is the position of atom $i$.  The
resulting value $D \approx 5.2 \times 10^{-9}$ m$^2$s$^{-1}$, combined
with the value $a = r_1 =2.65$~\AA, gives $\eta \approx 1.6 $ mPa~s,
which agrees quite well with the Green-Kubo value.
For the system at the experimental density we calculate $D \approx 6.8
\times 10^{-9}$ m$^2$s$^{-1}$, which is consistent with the value of
the viscosity.

Although the statistical errors on $\eta$ are small, one might ask
whether a 64-atoms system is big enough to calculate $\eta$
accurately. The effect of system size on calculated viscosities has
been extensively studied for models such as the hard-sphere and
Lennard-Jones liquids---close packed systems which are similar to
liquid aluminum. For example Schoen and Hoheisel \cite{schoel}
examined Green-Kubo calculations of the Lennard-Jones liquid for
system sizes from 32 to 2048 atoms. They showed that even 32 atoms
give meaningful values for $\eta$ and that for 64 atoms the error on
$\eta$ at the triple point is $ \approx 10 \%$.  We note that one
indicator of system size effects should be the difference between
$\eta$ values calculated with the stress components $P_{xy}$ etc. and
those calculated from $\frac{1}{2}(P_{yy}-P_{zz})$ etc., since cubic
periodicity breaks the spherical equivalence of the two types of SACF.
We searched for this difference in our calculations, but it is not
large enough to be clearly visible above the statistical noise.

Liquid iron under Earth's core conditions provides a good example of
the importance of being able to calculate transport coefficients by
FPMD. The Earth's liquid outer core consists mainly of molten iron,
with light impurities, of which sulfur is a likely candidate
\cite{sulfur}. The shear viscosity of this liquid is a crucial factor
in understanding convective circulation in the core, which is
intimately linked both to heat transport and to the generation of the
Earth's magnetic field. Nevertheless it is one of the most uncertain
quantities in geophysics, with estimates from different experimental
and theoretical methods spanning no less than 12 orders of magnitude
\cite{secco}.  A firmly based calculation of the viscosity is
therefore important. In our recent FPMD calculations for liquid Fe
\cite{vocadlo,nature} and liquid Fe/S \cite{prb} under Earth's core
conditions we used the Stokes-Einstein relation to obtain the estimate
$\eta = 13 $ mPa s. But the use of this relation under such extreme
conditions can clearly be challenged.

We present here the results that we have obtained from a direct
calculation of the viscosity of an Fe/S liquid alloy under Earth's
core conditions.  The calculations were performed at a thermodynamic
state corresponding to the boundary between the solid inner core and
the liquid outer core.  In fact, the temperature at this boundary is
uncertain, with estimates ranging from 4000 to 8000 K, while the
density is quite accurately known to be about $12000$ kg~m$^{-3}$
\cite{prem}.  The temperature in our simulation was set to 6000 K and
the density was 12330 kg~m$^{-3}$.  The Fe/S mixture was taken to have
a sulfur mole fraction of 0.1875, in line with the maximum estimates
for sulfur abundance in the core \cite{arhens}.  The simulated system
contained 64 atoms (52 Fe and 12 S) in the repeated cell, with
$\Gamma$ point sampling only.  We have used the generalized gradient
approximation \cite{gga} for the exchange-correlation energy, a plane
wave cutoff of 350 eV, a time step of 1 fs and a self-consistency
threshold of $1.5 \times 10^{-7}$ eV/atom. The total length of the
simulation was 10 ps and the drift on the total energy was $\approx 8$
meV/atom per ps.  A complete description of the results of this
simulation is reported elsewhere \cite{prb}.

\begin{figure}
\centerline{\psfig{figure=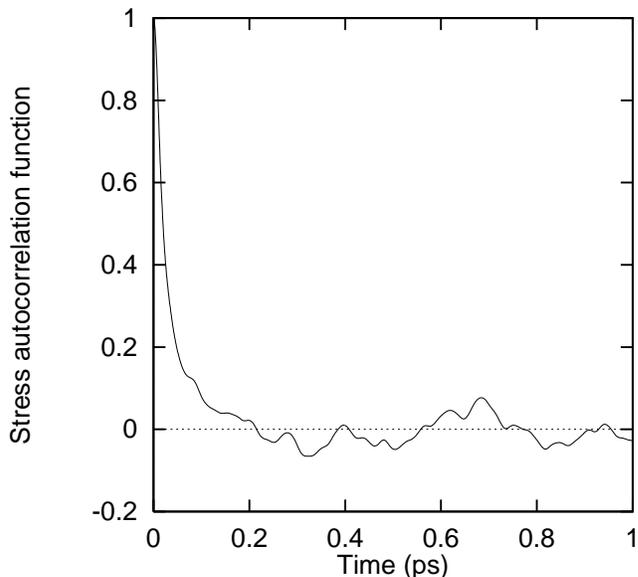,height=3in}}
\caption{Average over the five independent components of the
autocorrelation function of the traceless stress tensor $\phi(t)$
calculated for liquid Fe/S under Earth's core conditions. Values of
$\phi(t)$ are normalized by dividing by $\phi(0)$.}\label{stressfe}
\end{figure}

In Fig. \ref{stressfe} we display the average SACF $\phi(t)$ for this
system. This is noisier than for the aluminum system, the statistical
error at any time being $\approx 4 \%$ of $\phi(t=0)$, but it is still
possible to produce a useful estimate for the viscosity.  The
$\phi(t)$ has gone to zero after $t \approx 0.2$ ps (Fig. 3), and the
time integral of $\phi(t)$ up to this point gives $\eta = 9 \pm 2 $
mPa~s (Fig. 4), which confirms the approximate correctness of the
value estimated using the Stokes-Einstein relation. The error has been
estimated as in the aluminum case.  This value of $\eta$ is about 10
times larger than the viscosity of typical liquid metals at ambient
pressure. This result is important since it provides concrete support for
the approximation commonly made in magnetohydrodynamic models that the
outer core is an inviscid fluid \cite{glatz,merril} undergoing
small-circulation turbulent convection \cite{melchior} rather than
large-scale global circulation.

We have chosen the case of liquid Fe/S to illustrate the geophysical
importance of first-principles calculations of viscosity, but there
are many other areas where calculated viscosities would be
valuable. The viscosity of the highly compressed hydrogen/helium
mixtures in the deep interior of the giant planets is a case in point.
There have been recent attempts to deduce this viscosity from measured
gravitational anomalies for Jupiter, and input from first-principles
calculations obtained using the methods we have described would be
very useful. Insight into the percolation of molten minerals in the
Earth and the other terrestrial planets could also be gained by first
principles methods. In a completely different context, the viscous
flow of materials is crucial in a wide variety of industrial processes.

Although we have focused here on the shear viscosity, first-principles
calculations should also be feasible for other transport coefficients
like the bulk viscosity, the thermal conductivity and chemical
interdiffusion coefficients, all of which contribute to the
attenuation of sound in fluids. 

In conclusion, we have demonstrated the feasibility of calculating
transport coefficients such as viscosity from first principles, and we
have presented results for the viscosity of liquid aluminum which
agree well with experiment. We illustrated the importance of such
calculations by presenting results for the viscosity of liquid Fe/S
under Earth's core conditions. These results support recent
approximate estimates which indicate that the viscosity of the Earth's
outer core is in the region of 10 mPa s. This is far lower than the
values sometimes inferred from seismic and other measurements, and has
important implications for the understanding of circulation in the
core. Finally, we have pointed out that the techniques employed should
also be applicable to other transport coefficients.

We acknowledge the support of NERC under grant GST/O2/1454 and computer
time allocated by the High Performance Computing Initiative.


\begin{figure}
\centerline{\psfig{figure=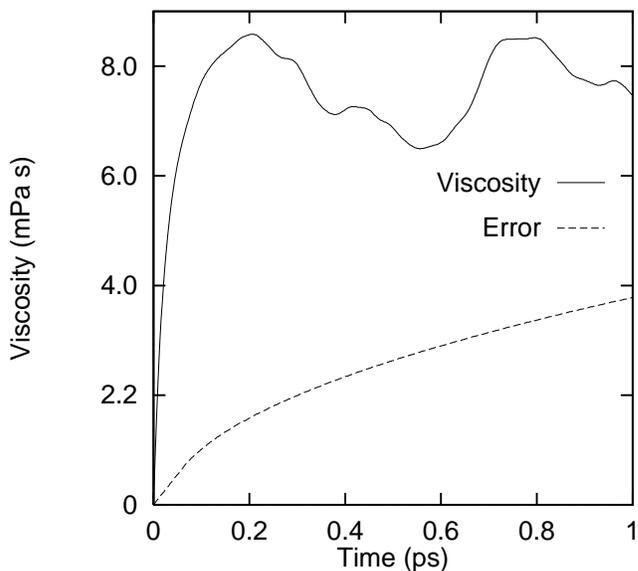,height=3in}}
\caption{Viscosity integral of the average stress autocorrelation
function and its statistical error as a function of time for liquid Fe/S
under Earth's core conditions.}\label{viscfe}
\end{figure}	

\end{document}